\documentclass[%
 reprint,
 superscriptaddress,
 showpacs,
  nofootinbib,
 amsmath,amssymb,
 aps,
 prc,
]{revtex4-1}

\usepackage{verbatim}
\usepackage{graphicx}
\usepackage{epsfig}
\usepackage{amsfonts}
\usepackage{amsmath,amssymb}
\usepackage{bm}
\usepackage{hyperref}

\begin{document}

\title{Ab-initio No-Core Gamow Shell Model calculations with realistic interactions}

\author{G. Papadimitriou}
\affiliation{
Department of Physics, University of Arizona, Tucson, AZ 85721, USA
}%

\author{J. Rotureau}
\affiliation{Fundamendal Physics, Chalmers University of Technology, 412 96 G\"{o}teborg, Sweden}

\author{N. Michel}
\affiliation{
Department of Physics and Astronomy, University of Tennessee, Knoxville, Tennessee 37996, USA
}%
\affiliation{
National Superconducting Cyclotron Laboratory and Department of Physics and Astronomy,
Michigan State University, East Lansing, MI 48824, USA
}%

\author{M. P{\l}oszajczak}
\affiliation{Grand Acc\'{e}l\'{e}rateur National d'Ions Lourds (GANIL), CEA/DSM-CNRS/IN2P3, BP 55027, F-14076 Caen Cedex 5, France}

\author{B. R. Barrett}
\affiliation{
Department of Physics, University of Arizona, Tucson, AZ 85721, USA
}%


\begin{abstract}
No-Core Gamow Shell Model (NCGSM) is applied for the first time to study selected well-bound and unbound states of helium isotopes. This model is formulated on the complex energy plane and, by using a complete Berggren ensemble, treats bound, resonant, and scattering states on equal footing. We use the Density Matrix Renormalization Group method
to solve the many-body Schr\"{o}dinger equation. To test the validity of our approach, we benchmarked the NCGSM results against Faddeev and Faddeev-Yakubovsky exact calculations for $^3$H and $^4$He nuclei. We also performed {\textit ab initio} NCGSM calculations for the unstable nucleus $^5$He and determined the ground state energy and decay width, starting from a realistic N$^3$LO chiral interaction. 
\end{abstract}


\maketitle



\section{Introduction}

In the last decade our knowledge of nuclei far from the valley of stability has radically improved.
This improvement is a by-product of advances in both experiment and theory.
New experimental facilities that have already been built (RIBF at RIKEN)
or are being constructed (SPIRAL2 at GANIL, FAIR, FRIB at MSU) will give us a better insight
of areas in the nuclear chart that have never been explored, pushing even farther the limits of
nuclear existence. A few decades ago,  the nuclear chart consisted of approximately 1000 isotopes,
whereas in 2011 this number has been expanded to approximately 3000 species, and an
estimated number of nuclei that can exist in nature or synthesized in the
laboratory is approximately 7000 \cite{Witek_nature}.
The increase in computing power has made it possible to calculate the properties
of nuclei in an \textit{ab initio} manner, using realistic interactions, which reproduce the nucleon-nucleon 
scattering data. For few-body systems (A $\leq$ 4) methods such as Faddeev \cite{nogga_fad} and Faddeev-Yakubovsky (FY) \cite{Yakubo}
provide an exact solution to the many-body problem.  Methods such as the Green's Function Monte Carlo (GFMC) \cite{gfmc}, the Hyperspherical Harmonics (HH) \cite{HH1,HH2}, the No-Core Shell Model (NCSM) \cite{peter,caprio}, the Coupled-Cluster approach (CC) \cite{cc_review}, and more recently, the In-Medium Similarity Renormalization Group method (IM-SRG) \cite{tsukiyama,hergert} and Dyson Self-Consistent Green's Function (Dyson-SCGF) method  \cite{vittorio} have been applied successfully for the \textit{ab initio} description of light and medium mass nuclei.
 
 Nuclei with a large isospin that can be found in regions far away from the valley of stability have attracted a great deal of interest.
 They belong to the category of  Open Quantum Systems (OQS) \cite{OQSbook1}, which in the case of the nucleus are inter-connected via the decay and reaction channels. They are very fragile objects with small separation energies and very large spatial dimensions. The proximity of the continuum affects  their bulk properties (matter and charge distributions) and  their spectra.
 Phenomena such as the anomalous behavior of elastic cross-sections and the associated overlap integral near threshold states in multi-channel coupling (Wigner-cusps) \cite{wigner_cusp,rf:18a}, the isospin and mirror symmetry-breaking threshold effects \cite{ehrman,GSM_isospin_break}, the resonance trapping \cite{rf:12,rf:13,rf:14} and super-radiance phenomenon \cite{rf:15,rf:16}, the appearance of cluster correlations in the vicinity of the respective cluster emission threshold \cite{opn1}, the modification of spectral fluctuations \cite{rf:19,rf:19a,rf:19b}, and deviations from Porter-Thomas resonance width distribution \cite{rf:14,rf:20,rf:20a},
 are all unique manifestations of the continuum coupling. 
 
For their theoretical explanation it was necessary to generalize existing many-body methods, and create theories which unify structure and reactions. Examples of these attempts are the  Shell Model Embedded in the Continuum (SMEC)  \cite{rf:7,Okolowicz,rot_smec} and the Gamow Shell Model (GSM) \cite{GSM_first_two,rf:2,review_GSM,gp}. The SMEC is a recent realization of the real-energy Continuum Shell Model \cite{rf:6,rf:9} which uses the Feshbach projection technique \cite{rf:10} in order to take into account the coupling to the scattering continuum. The GSM is a generalization of the Harmonic Oscillator based shell model in the complex energy plane by using the Berggren ensemble \cite{Berggren}, which treats resonant and non-resonant states on equal footing.
 \textit{Ab initio} calculations that can describe bound and unbound states of nuclei, 
 include the NCSM coupled with the Resonating Group Method \cite{quaglioni,baroni,hupin}, the CC approach generalized in the complex-energy plane using the Berggren basis \cite{Gaute_halo,hagen_f17,hagen_oxygen_ca48} and the GFMC \cite{Nollett,Nollett_12}.

 In this work we introduce the No-Core Gamow Shell Model (NCGSM) as an alternative for calculations of weakly bound and
 unbound states of light nuclei using realistic interactions and allowing all the nucleons to be active.
 The paper is organized in the following  manner:
 in Sections \ref{hamiltonian} and \ref{Basis_tbmes_etc}, we describe the basic ingredients of our method, such as the many-body Hamiltonian, the single-particle basis we employ,
 the way the two-body matrix elements are calculated within the Berggren basis, and we discuss the translational invariance
 of our Hamiltonian. In Section \ref{dmrg} we describe the Density Matrix Renormalization Group (DMRG) method, which
 is an efficient tool for a diagonalization of large complex-symmetric GSM matrices. In Section \ref{results} we present our calculations
 for the $^3$H, $^4$He and $^5$He nuclei and, finally, in Section \ref{conclusions} we discuss the conclusions and the
 future perspectives.

\section{Hamiltonian}
\label{hamiltonian}

Our goal is to solve the $A$-body Schr\"{o}dinger equation 
\begin{equation}
H|\Psi\rangle = E|\Psi\rangle \label{many_eq} ,
\end{equation}
where $H$ is the intrinsic Hamiltonian
\begin{equation}
\mathit{H} = \frac{1}{A}\sum_{i<j}^{A}\frac{ \left( \vec{p}_{i} - \vec{p}_{j} \right)^{2} }{2m} + \sum_{i<j}^{A}V^{NN}_{ij},
\label{GSM_hami}
\end{equation}
and $V^{NN}$ a realistic NN interaction. For  \eqref{GSM_hami} the following identities are useful:
\begin{figure}[h!] 
  \includegraphics[width=\columnwidth]{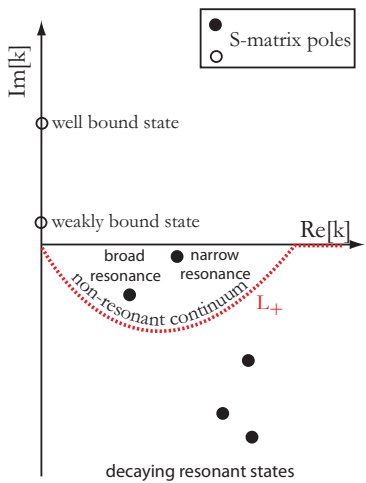}
  \caption[T]{\label{fig1}
  (Color online)     An illustration of the Berggren s.p. basis used in the NCGSM, showing the position of resonant (bound states and resonances) states in the complex $k$-plane. The non-resonant continuum states lie along the complex contour $L_{+}$.  }
\end{figure}
\begin{equation}
\vec{P} = \sum_{i=1}^{A} \vec{p_{i}},
\end{equation}
where $\vec{P}$ is the Center of Mass (CoM) momentum,
and
\begin{equation}
\sum_{i=1}^{A} \vec{p}_i^2 = \frac{1}{A}\left[ \vec{P}^2 + \sum_{i<j} \left( \vec{p_i} - \vec{p_j} \right)^2  \right],
\end{equation}
resulting in:
\begin{equation}
\sum_{i=1}^A \frac{\vec{p}_i^2}{2m} - \frac{\vec{P^2}}{2mA} = \frac{1}{2mA}\sum_{i<j} \left( \vec{p_i} - \vec{p_j} \right)^2
\end{equation}
In the NCGSM there is no restriction on the type of the NN interaction, contrary to the GFMC approach for example, where difficulties arise for non-local potentials. One can use a local interaction, such as the Argonne $\upsilon_{18}$ potential \cite{Argonne_v18}, or a non-local interaction, such as the CD-Bonn 2000 \cite{CD_Bonn} and various chiral  interactions. There is also the possibility to use renormalized versions of the aforementioned forces, by applying techniques such as $V_{low-k}$ \cite{Scott}, the Similarity Renormalization Group (SRG) approach \cite{Jurgenson,Roth}, or the G-matrix \cite{morten_prp,cens}. In this work we employ the phenomenological Argonne $\upsilon_{18}$ potential and the chiral N$^3$LO interaction \cite{n3lo_pot} which is consistent with the symmetries of the QCD Lagrangian.  Both potentials were renormalized via the $V_{low-k}$ method with a sharp momentum cutoff $\Lambda$ = 1.9 fm$^{-1}$ to decouple  high from low momentum degrees of freedom and, henceforth, improving the convergence of nuclear structure calculations \cite{vlowk_improv}. Moreover, specific interactions will be used to compare NCGSM with other approaches.

\section{Berggren basis}
\label{Basis_tbmes_etc}

In previous applications of the GSM, where a tightly bound core was assumed ($^4$He or $^{16}$O), the single particle (s.p.) basis  was usually generated by solving the one-body Schr\"{o}dinger equation with a Woods-Saxon (WS) potential,
parameterized to reproduce the core plus nucleon spectrum. In the case of the NCGSM,
the s.p. basis will be generated by the realistic two-body interaction itself by solving the integro-differential Schr\"{o}dinger equation  which contains both local and non-local parts \cite{GSM_GHF,veneroni}. This numerical method is known as the Gamow Hartree-Fock (GHF) method since it generates a microscopic basis that includes resonant and non-resonant states.  The GHF method can be applied not only in spherical cases (closed-shells) but also in deformed cases (non-closed shells) \cite{GSM_GHF}. 
The one-body self-consistent potential U$_{HF}$(r) is then used to solve the one-body Schr\"{o}dinger equation:
\begin{equation}
u^{\prime \prime}_{k}(r) = \left [   \frac{\ell(\ell + 1)}{r^2} + \frac{2m}{\hbar^2}U_{HF}(r) + V^{c}(Z_c,r) - k^2   \right ]u_{k}(r)
\label{one_body_hami}
\end{equation}
where $V^c(Z_c,r)$ is the one-body Coulomb potential:
\begin{equation}
V^c(Z_c,r) = \frac{C_c Z_c \mathrm{erf}(\alpha r)}{r}
\label{Coulomb_pot}
\end{equation}
and  $C_c$ is the Coulomb constant, $Z_c$ the proton number and $\alpha$ is a constant, which is given by $\alpha$ = $3\sqrt{\pi}/4R_{0}$. The reason we choose an error function to approximate the Coulomb field and not for example, the field that is produced by a uniformly charged sphere at $R_{0}$, lies in the fact that the latter is non-analytic at $R_{0}$. The value of $R_{0}$ is chosen in a way that the Coulomb potential of Eq. \eqref{Coulomb_pot} and the potential of a uniformly charged sphere are equal at the origin. 
The wave number $k$ is defined as: $k$ = $\sqrt{2mE}/\hbar$ and is, in general, complex.
Equation \eqref{one_body_hami} is solved with the requirement that at large distances the wave function will behave as a linear combination of Hankel or Coulomb functions for neutrons and protons, respectively:
\begin{equation}
u_{k}(r) \sim C^{+}H^{(+)}_{\ell, \eta}(kr) + C^{-}H^{(-)}_{\ell,\eta}(kr)
\label{Hankel_coef}
\end{equation}
where $\eta$ is the Sommerfeld parameter.
The $C^+$ and $C^-$ coefficients are determined by the normalization of the radial functions $u(r)$ to a Dirac's $\delta$ distribution:
\begin{equation}
\int_{0}^{\infty}u_{k}(r)u_{k^{\prime}}(r)\,dr = \delta(k-k^{\prime})
\end{equation}
\begin{figure}[h!] 
  \includegraphics[width=\columnwidth]{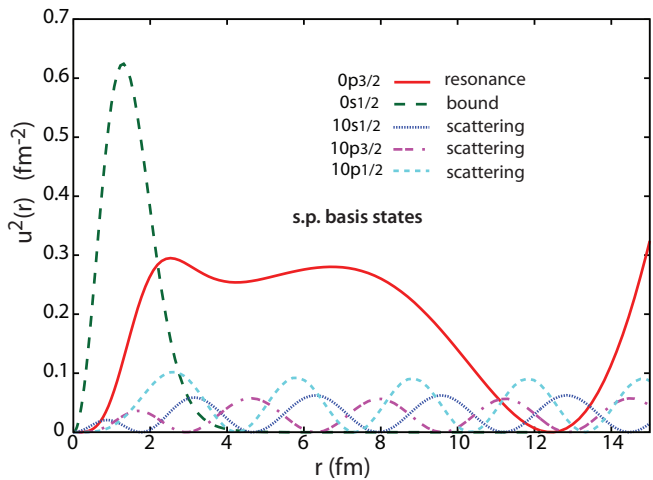}
  \caption[T]{\label{fig2}
  (Color online) Selected squared radial basis functions in the NCGSM. Resonant (bound, resonances) and non-resonant (scattering) states are plotted. The basis generating potential is produced by the $V_{low-k}$ N$^3$LO interaction for the $^5$He nucleus. The notation $nlj$ is explained in the text.}
\end{figure}
The solutions of \eqref{one_body_hami} which satisfy pure outgoing boundary conditions ($C^-$ = 0 in \eqref{Hankel_coef})  
correspond to the poles of the S-matrix and they are represented as dots in the complex $k$-plane of Fig. \ref{fig1}.
While normalization of bound  states does not pose any difficulty, one should pay more attention on the normalization of resonances.
The latter  diverge exponentially for large distances and the regularization method that is used for the calculation of their norm is the external complex scaling \cite{review_GSM}.
The method facilitates from the fact  of using complex radii for the integration of resonant wave functions: 
\begin{equation}
\int_{0}^{\infty} u_{k}^2(r) \, dr  = \int_{0}^{R} u_{k}^2(r) \, dr + \int_{R}^{\infty} u_{k}(R+xe^{i\theta})^2e^{i\theta} \, dx,
\end{equation}
with $R$ chosen sufficiently large so as to match the condition \eqref{Hankel_coef} for $C^{-}$ = 0, while $\theta$ is the angle of
external rotation, which satisfies the condition that $u_{k}(R + xe^{i\theta})$ = 0 for $x \to \infty$.

It was shown by Berggren \cite{Berggren} that for a given partial wave ($\ell$,$j$) the scattering states, which are distributed along the $L_{+}$ contour and the resonant solutions (bound states and/or resonances) of \eqref{one_body_hami} form a complete set:
\begin{equation}
\sum_{n}| u_{n} \rangle \langle \widetilde{u_{n}} | + \int_{L_+} |u_k \rangle \langle \widetilde{u_k} | dk = 1.
\label{complet}
\end{equation}
The tilde symbol means that the complex conjugation arising in the dual space affects only the angular part and leaves the radial part of the wave function unchanged. The completeness reassures us that any function which lies between the contour and real $k$-axis and exhibits outgoing wave asymptotic ($e^{ikr}$), can be expanded using \eqref{complet}.  In practice the integral in \eqref{complet} is discretized by means of an appropriate quadrature rule (the Gauss-Legendre quadrature in our case) and we end up with a discretized completeness relation:
\begin{equation}
\sum_{i=1}^{N}\omega_{n}|u_n\rangle \langle \widetilde{u_n}| \simeq 1,  
\label{compl_disc}
\end{equation}
where $\omega_n$ = 1 both for resonant states and the Gauss-Legendre weight for non-resonant states along the discretized contour.
The approximate equality in \eqref{compl_disc} arises from the finite discretization of the contour.
In addition, the discretized contour does not extent to infinity and we use a maximum cutoff ($k_{max}$) at around 4 fm$^{-1}$.
We have checked that results do not depend on the values of the $k_{max}$ for an adequately large number of discretization points.

In Fig. \ref{fig2} we show the radial behavior of a few Berggren basis  states that were generated from the N$^3$LO interaction with 
a $V_{low-k}$ cut-off $\Lambda$ = 1.9 fm$^{-1}$. The states that are plotted refer to neutron states.
The 0$p_{3/2}$ resonant state has a complex energy and  lies in the fourth quadrant of the complex $k$-plane (see also Fig. \ref{fig1}). It
is a solution of Eq. \eqref{one_body_hami} with outgoing wave boundary conditions at large distances. We, indeed, observe a 
localization of the state in the region of the attractive nuclear potential (r $\leq$ 2 fm) and an outgoing wave behavior for distances beyond the range of the nuclear potential. This is the basic characteristic of a metastable s.p. state.  

In order to satisfy Berggren's completeness, the scattering contour $L_{+}$ has to be complex. It is then understood that the 10$p_{3/2}$ is a point along the complex contour. It corresponds to a state, which is given as a linear combination
of Hankel functions, as it is seen in Eq. \eqref{Hankel_coef}.  Here we are using the notation $n$$\ell$$j$, where $n$ is the radial quantum number and is identified as the 10th Gauss-Legendre discretization point on the $L_{+}$ contour; $\ell$ is the s.p. angular momentum of the state ($\ell$ = 1 in this example), and $j$ is the s.p. total angular momentum. 
The 0$s_{1/2}$ resonant state is bound and lies on the imaginary momentum axis with a real and negative energy. At large distances, the
state is decaying as an exponential. Also shown are the 10$s_{1/2}$ and 10$p_{1/2}$ scattering states,
which both lie on the real momentum axis, since there is no complex resonant state associated with these partial waves. Similar to the 10$p_{3/2}$ complex state, they correspond to the 10th discretization point of the Gauss-Legendre rule, but the contour lies along the real momentum axis. At this point we would like to highlight that the Berggren basis not only imposes the correct quantum mechanical asymptotic behavior of the s.p. states \footnote{This can be immediately checked by the fact that a bound resonant state ($\mathrm{e^{ikr}}$) behaves as  $\mathrm{e^{-\kappa r}}$ for $r\to\infty$, since $k = i\kappa$. } but also includes the continuum states in a rigorous manner, promoting it to an ideal and realistic basis for the description of metastable and weakly bound states.

The completeness relation \eqref{compl_disc} is similar to a usual discrete completeness relation, such as the Harmonic Oscillator (HO)) one, and results in an eigenvalue problem. The many-body basis states are Slater determinants $|SD_n\rangle=|u_1\ldots u_A\rangle$, where $u_k$ is a resonant (bound state or resonance) or non-resonant (scattering) state. 
In this basis, the Hamiltonian matrix is complex symmetric and upon diagonalization, many-body correlations
and coupling to the continuum are taken into account simultaneously. 
As a direct consequence of Eq. \eqref{compl_disc}, the  many-body states also satisfy the completeness relation:
\begin{equation}
\sum_{i=1}^{N}|\widetilde{SD_n}\rangle \langle SD_n| \simeq 1. 
\label{compl_many}
\end{equation}
The squares of the linear expansion coefficients and not by their absolute values, satisfy the relation:
\begin{equation}
\sum_{i=1}^{N} c_{n}^2 = 1.  
\label{compl_ampli}
\end{equation}
Furthermore, the completeness relation \eqref{compl_many} can be used for the calculation of two-body matrix elements (TBMEs)
between Berggren basis states.

\subsection{TBMEs of realistic interactions in the Berggren basis}

---\textit{Nuclear part}---
Matrix elements of realistic interactions are defined in a relative and CoM system of coordinates. In order to work in a basis of Slater determinants, a transformation from the relative and CoM to the laboratory system is necessary. When working in the HO s.p. basis, this is possible through the Brody-Moshinsky brackets \cite{Mosh1}. For a different basis such as the Berggren basis, one has to perform a multiple decomposition of the realistic NN interaction and calculate two-dimensional radial integrals. Matrix elements between scattering states need to be regularized by means of the complex scaling (CS) technique \cite{review_GSM} which unfortunately does not work for just any type of integrals and can cause numerical instabilities.
The problem is alleviated by expanding the NN interaction in a truncated HO basis \cite{hagen_morten_michel}:
\begin{equation}
V_{NN} = \sum_{\alpha \beta \gamma \delta}^{N_{max}} |\alpha \beta \rangle \langle \alpha \beta|V_{NN}| \gamma \delta \rangle \langle \gamma \delta |
\label{HO_exp}
\end{equation}
The matrix elements of the NN interaction in the Berggren ensemble become then:
\begin{equation}
\langle \widetilde{ab}| V_{NN} | cd \rangle = \sum_{\alpha \beta \gamma \delta}^{N_{max}} \langle \widetilde{ab}|\alpha \beta  \rangle \langle \alpha \beta|V_{NN}| \gamma \delta \rangle \langle \gamma \delta | cd \rangle
\end{equation}
We end up calculating overlaps between HO and Berggren states 
$\langle \alpha \beta | a b \rangle$ and $\langle \gamma \delta | c d \rangle$,
where the Latin letters denote Berggren states and Greek letters HO states. Due to the Gaussian fall-off of the HO states, no complex scaling is needed for the calculation of these integrals. On the other hand, matrix elements of the NN interaction in the HO basis $\langle \alpha \beta|V_{NN}| \gamma \delta \rangle$ can be conveniently calculated using the Brody-Moshinsky brackets \cite{Mosh1}. 

The method of handling matrix elements in the Berggren basis using the projection of continuum onto the HO states should not be confused with basis expansion methods suitable for the description of closed quantum systems. Only the short range part of the nuclear interaction is expanded in the HO basis. The kinetic energy operator is calculated in the Berggren basis, so the calculations of weakly bound and unbound systems are possible. With this formulation it is also clear that there is no restriction on the type of NN interaction one can use in the NCGSM. What we need is just the nuclear matrix elements in the HO basis.

---\textit{Coulomb interaction treatment}---
For nuclear systems with two or more protons the two-body Coulomb interaction is included in the Hamiltonian \eqref{GSM_hami}.
The method we adopt for the treatment of the long-range Coulomb interaction was first used in the description of isospin breaking due the continuum coupling \cite{GSM_isospin_break} and it was also recently applied to calculate reaction observables \cite{yan02,gaute_michel_reactions} (see also \cite{nicolas_coulomb} for a detailed numerical analysis).
The basic idea of the method is to add and subtract the one-body Coulomb potential \eqref{Coulomb_pot} (with $Z=2$, e.g., in case of $^3$He or $^5$He) from the two-body Coulomb interaction: $V_c(1,2) = V^c(1) + \left( V_c(1,2) - V^c(1) \right)$. Then the second term in the parentheses has a short-range character and the HO expansion method of \eqref{HO_exp} can be applied.
Matrix elements of the Coulomb interaction can be calculated using the Brody-Moshinsky brackets without the need to perform an external CS calculation.

We would like to mention here that this method of treating the two-body Coulomb interaction would be of particular interest, when one has to deal with many-body proton resonances, such as in the $^6$Be nucleus. In this case, calculating the Coulomb interaction potential by simply expanding it in a HO basis (see Eq. \eqref{HO_exp}), would be a rather poor approximation.

---\textit{Center of mass (CoM) motion in the NCGSM}---
 By adopting a s.p. basis upon which we build many-body basis states, we effectively localize the nucleus in space and, hence, we break the translational invariance of the Hamiltonian. Moreover, we would need $3A-3$ coordinates to describe it,
but the nuclear wave function we construct, depends on $3A$ coordinates, where $A$ is the total number of particles. These redundant degrees of freedom are responsible for the CoM spuriosity that appears in many-body methods. On the other hand, plane waves s.p. states are eigenstates of the momentum operator and, hence, preserve the translational invariance, but unfortunately cannot be used to  describe a localized system. The alternatives are: i) Solving the many-body problem using relative coordinates (e.g. Jacobi coordinates) which reassures the translational invariance of the system, with the price of unfeasible antisymmetrization of states for $A>$8, and ii) Using the unique analytical properties of the HO s.p. basis in a full $N\hbar\omega$ space, in which the total wave function
is factorized into $|\psi_{\mathrm{rel}}\rangle$$\otimes$$|\psi_{\mathrm{CoM}}\rangle$, limiting though the application to well-bound systems only. In the case of the NCGSM the latter factorization is not guaranteed and it has to be demonstrated numerically. 
Since our Hamiltonian \eqref{GSM_hami} is intrinsic, we are expecting that in an infinite space
there would be no spuriosity. However, because we are working in a finite space, it is necessary to check numerically this condition. 

Assuming that the factorization into a CoM and a relative wave function is valid and also  the CoM wave function has a Gaussian shape, we calculate the expectation value of the CoM operator \cite{Palumbo}: 
\begin{equation}
H_{CoM} = \frac{1}{2mA}\vec{P}^2_{cm} + \frac{mA\omega^2}{2}\vec{R}^2_{cm} - \frac{3}{2}\hbar\omega,
\label{Hcm_op}
\end{equation}
where $\hbar\omega$ is the parameter that characterizes the Gaussian wave function.
The matrix elements of \eqref{Hcm_op} are calculated with the HO expansion method of \eqref{HO_exp} and the analytical
formulas for their expression are found in \cite{Lawson}. Following the assertion of  Ref. \cite{Gaute_Thomas_CoM}, if $\langle H_{cm} \rangle \sim $ 0 then the factorization is valid.

\section{Resolution of the many-body Schr\"{o}dinger equation with the DMRG method}
\label{dmrg}

The Schr\"{o}dinger equation \eqref{many_eq} is solved within a many-body basis constructed from the discrete set of single-particle
states $|u_i\rangle$ \eqref{compl_disc}. The discretization of the integral along the contour $L_+$ in Eq. \eqref{complet} should be precise enough so that the discretized completeness relation \eqref{compl_disc} is fulfilled. In other words, the number of discretized shells
 should be increased until Eq. \eqref{compl_disc} holds. As a consequence, the dimension of the many-body  model space will increase dramatically with the number of nucleons and number of shells. Efficient numerical methods allowing the diagonalization of large Hermitian as well as complex-symmetric matrices are then required to solve the NCGSM problem.

In this paper, we have used one of these methods, namely, the DMRG method \cite{dmrg1,pap6} which has been generalized in the context of the GSM in \cite{dmrg_rot1,dmrg_rot2}. The GSM/DMRG approach has been applied previously to study several weakly-bound/unbound nuclei described as few-valence-nucleon systems interacting via schematic two-body forces above an inert core. In this paper, all nucleons are considered active and realistic two-body interactions are used but nevertheless, the application of the DMRG method is similar. In the following, we recall the main ideas of the DMRG in the multi-shell GSM problem \cite{dmrg_rot2}.

The purpose of the DMRG method is to allow the calculation of the many-body poles of the scattering matrix of the NCGSM Hamiltonian $\hat{H}$ by performing efficient truncations of the many-body model space. As the contribution of the non-resonant continuum to the structure of many-body bound/resonant eigenstates of  $\hat{H}$ is usually smaller than the contribution from the bound/resonant orbits, the following separation is performed: the many-body states constructed from the s.p. poles form  a subspace $H$ 
(the so-called 'reference subspace'), and the remaining states containing contributions from  non-resonant  shells form a complement subspace $P$.
The set ${\cal E}$ of many-body basis states (\ref{compl_many}) can then be written as 
\begin{equation}
{\cal E}=H\otimes P.
\end{equation}
The DMRG technique is then used to perform truncations in $\cal{E}$ by keeping only selected 'optimized' states in $P$ in the sense of a criteria based on the density matrix in $P$ (see below for a more explicit explanation). 

Let us assume that we want to calculate an eigenstate $|\Psi\rangle$ of $\hat{H}$ 
for a nucleus coupled to the total angular momentum $J$ and parity $\pi$. The number of proton(s) and neutron(s) are respectively $n_{\pi}$ and $n_{\nu}$. One begins by constructing all states $|k\rangle_H$ forming the  subspace $H$. The set of those states is
denoted  as $\{ k_H\}$. The many-body configurations in $H$ can be classified in different families  $\{n;j_H^\pi \}$
according to their number of nucleons $n$, total angular momentum $j_H$, and parity $\pi$. States with a number of protons (neutrons) larger than $n_{\pi}$ ($n_{\nu}$) are not considered since they do not contribute to the many-body states in the composition
of subspaces $H$ and $P$. The matrix elements in $H$ of the suboperators of the NCGSM Hamiltonian $\hat{H}$ 
expressed in the second quantization form, are calculated and stored: 
\begin{eqnarray}\label{suboper}
 \{O\}& =& \{a^{\dagger },
(a^{\dagger }\, \widetilde{a})^{K}, (a^{\dagger }a^{\dagger
})^{K}, ((a^{\dagger }a^{\dagger })^{K} \widetilde{a})^{L}, \nonumber \\
&& (a^{\dagger }a^{\dagger })^{K}
(\widetilde{a}\widetilde{a})^{K}\},
\end{eqnarray}
where $a^{\dagger }$ and $\widetilde{a}$  are the nucleon creation and annihilation operators in shells forming the subspace $H$.
The NCGSM Hamiltonian is then diagonalized in the pole space $H$ to provide the zeroth-order  approximation
$|\Psi\rangle^{(0)}$ to $|\Psi\rangle$. 

In the following stage, the subspace $P$ is built, step by step, by adding scattering shells one by one during the so-called  'warm-up phase'. At each step, many-body states constructed within the new added shell are coupled to optimized many-body states constructed 
 during previous steps {\i.e.}, constructed within previously added shells. Moreover, the matrix elements (\ref{suboper}) of the suboperators acting among the optimized states have been stored during previous steps.

\begin{figure}[h!]
\begin{center}
\includegraphics[width=0.35\textwidth]{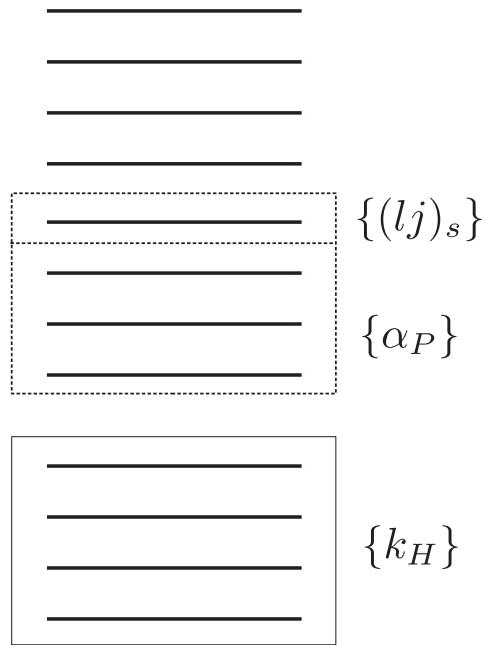}
\caption{Schematic illustration of the NCGSM/DMRG procedure  during the $s^{th}$ step of the warm-up phase.
States $\{k_{H} \}$ from $H$, previously optimized states $\alpha_{H} $, and states $\{(lj)_s \}$ constructed by occupying  the $s^{th}$ shell with $n$ particles are coupled to generate the new set of states $\{k_H \otimes i_P\}^J=\{k_{H} \otimes  \{\alpha_P \otimes (lj)_s^n \} \}^J $.}
\label{fig3}
\end{center}
\end{figure}

To be more specific, let us assume that the $s^{th}$ step is reached. The method is illustrated in Fig. \ref{fig3}.
The scattering shell $(lj)_s$  belonging to the discretized contour $L^{+}$ is added and within this shell, one constructs all possible many-body states $\{(lj)^{n_P}_s\}$. Matrix elements of suboperators (\ref{suboper})  acting on  $\{(lj)^{n_P}_s\}$ are also computed. One then couples previously optimized states denoted as $|\alpha\rangle_P$ to $\{(lj)^{n_P}_s\}$ to obtain the set of states $\{i_P\}=\{\alpha_P \otimes (lj)^{n_P}_s\}$.  States in $H$  are then coupled with the states $|i_P\rangle$ to construct the set
$\{ k_H \otimes  i_P \}^{J}$ of states coupled to $J^{\pi}$  that constitutes a basis in which the NCGSM Hamiltonian is diagonalized. The Hamiltonian matrix is constructed in this set with the Wigner-Eckart theorem and the matrix elements of the suboperators \eqref{suboper} acting on $\{k_H\}$,$\{\alpha_P\}$ and $\{(lj)^{n_P}_s\}$.

The target state $|\Psi\rangle$ is  selected among the eigenstates of $\hat{H}$ as the one having the largest overlap with the reference vector $|\Psi\rangle^{(0)}$. Based on the expansion
\begin{equation}
|\Psi\rangle = \sum_{k_H,i_P} c^{k_H(j_H)}_{i_P(j_P)} \{ |k_H(j_H)\rangle \otimes  |i_P(j_P)\rangle \}^J, \label{psi_exp}
\end{equation}
by summing over the  reference subspace $H$ for a {\em fixed} value of $j_P$,
one defines the reduced density matrix \cite{Mcdmrg}:
\begin{eqnarray}\label{rdm}
\rho ^{P}_{i_Pi'_P}(j_P) \equiv \sum_{k_H}c^{k_H(j_H)}_{i_P(j_P)} c^{k_H(j_H)}_{i'_P(j_P)}.
\end{eqnarray}

Truncation in the subspace $P$ is dictated by  the density matrix. In the case of a Hermitian Hamiltonian, the eigenvalues of the density matrix  are real and one can show that the truncation in $P$ is optimal when one keeps the eigenstates of the density matrix with the largest eigenvalues \cite{white_vp}. More specifically, the error in the representation of $|\Psi\rangle$ (\ref{psi_exp}) {\it after truncation}, is minimal in that case. In that sense, the eigenstates of the density matrix are 'optimal'.

Within the metric defining the Berggren ensemble, the NCGSM density matrix is complex-symmetric and its eigenvalues  are, in general, complex. The trace of the density matrix being equal to one, the truncation is done by keeping eigenstates of the density matrix with the corresponding eigenvalue $w_\alpha$ such that the condition
\begin{eqnarray}
\left| 1-Re\left(\sum_{\alpha=1}^{N_{\rho}}w_\alpha\right)\right|< \epsilon
\label{criterion_den}
\end{eqnarray}
is satisfied. The quantity $\epsilon$ in \eqref{criterion_den} can be viewed as the truncation error of the reduced density matrix.  The smaller $\epsilon$,  the larger number of eigenvectors must be kept. In particular, for  $\epsilon$=0,  all eigenvectors with non-zero eigenvalues are retained.

One then keeps eigenstates of the density matrix according to Eq. (\ref{criterion_den}). These are expressed as linear combination of the vectors $|i\rangle_P$ in $P$ and all matrix elements of the suboperators in these optimized states are recalculated and stored. Note that at each step, we enforce that at least one state in each family $\{n;j_P^\pi \}$ is kept \cite{dmrg_rot2}.

The warm-up phase continues by having the $P$ subspace grow by adding scattering shells one by one until the last shell is reached, providing a first guess for the wave function of the system in the whole ensemble of shells. At this point, all s.p. states have been considered, and all suboperators of the Hamiltonian $\hat{H}$ acting on states saved after truncation in $P$ have been computed and stored. The warm-up phase ends  and the so-called sweeping phase begins. 

Starting from the last scattering shell $(lj)_{last}$, the procedure continues in the reverse direction (the 'sweep-down' phase) using the
previously stored information. At this stage, the truncations are done according to the density matrix, which is obtained by summing over states $|k_H\rangle$ of the reference subspace $H$ and the states $|{i_{prev}\rangle}$ generated in the warm-up phase. Scattering
shells are added one at a time and at the last step of the sweep down phase, the first scattering shell is reached. The procedure is then reversed and a sweep in the upward direction (the 'sweep-up' phase) begins. Using the information previously stored, a first shell is added, then a second one, etc. The sweeping sequences continue until convergence for the target eigenvalue is achieved. For more details, see Ref. \cite{dmrg_rot2}.

\section{Applications of the NCGSM/DMRG}
\label{results}

\subsection{Test of convergence with respect to $N_{max}$ of the two-body interaction in HO expansion}

In Table \ref{tab:1} we check the energy convergence of the $^3$He nucleus with respect of the $N_{max}$ parameter in the expansion of the two-body interaction. We choose $^3$He due to the existence of both nuclear and Coulomb parts in the NCGSM Hamiltonian. The interaction employed is the N$^3$LO renormalized at a cut-off $\Lambda$ = 1.9 fm$^{-1}$, and we used a s.p. Berggren basis consisting only of $s_{1/2}$, $p_{1/2}$ and $p_{3/2}$ orbitals, for both neutrons and protons. 
\begin{table}[ht]
\caption{Convergence of the NCGSM eigenvalues with respect to the $N_{max}  = 2n_{max} + \ell_{max}$ of the HO expansion of the NN interaction. Values are in MeV. The length parameter of HO states is $b = 1.5$ fm. For this value of the $b$ parameter, the CoM wave function is a Gaussian. }
\begin{center}
\begin{tabular}{cc}
  \hline
   N$_{max}$& Energy \\
   \hline
  5 &  -5.321\\
  7 &  -5.334  \\
  9 &   -5.336 \\
  11 &  -5.343\\
  13 &      -5.343 \\
  15  &  -5.343 \\
  17   &   -5.346 \\
  19   &   -5.349 \\
  21  &  -5.352 \\
  23  & -5.352 \\
  25  & -5.352 \\
  27  &  -5.352 \\
  29  &  -5.352 \\
\end{tabular}
\label{tab:1}
\end{center}
\end{table}
The 0$s_{1/2}$ states are bound and the scattering contours lie on the real momentum axis. They are discretized with 20 points each and they extend up to 4 fm$^{-1}$. At this point, our purpose is not to describe realistically $^3$He but rather check the convergence of the many-body result with respect to a number of HO states of the expansion of the realistic interaction. In Table \ref{tab:1} we see that there is an overall small variation of the energy with increasing $n_{max}$ (or $N_{max}$). The energy changes only by $\sim 30$ keV by changing $N_{max}$ from $N_{max}$ = 5 to $N_{max}$ = 29. For $N_{max}$ = 21 ($n_{max}$ = 10), the energy shows the convergence and in all the following we will adopt $n_{max}$ = 10 for our calculations.

\subsection{Convergence with respect to the angular momentum of the model space and with respect to 
the number of sweeps in the DMRG method}

In this section, we are applying the NCGSM for $^3$H nucleus to test the DMRG method for solving the many-body Schr\"{o}dinger equation in Berggren basis. $^3$H (and also $^4$He or $^3$He) is a well bound system  so results of the NCGSM/DMRG approach can be compared against other well-known bound-states methods. The numerical task in solving a three nucleon system is relatively easy so that one can  check the convergence of NCGSM/DMRG results with respect to the maximal angular momentum of the model space and the truncation error of the reduced density matrix. 

\begin{figure}[h!] 
  \includegraphics[width=\columnwidth]{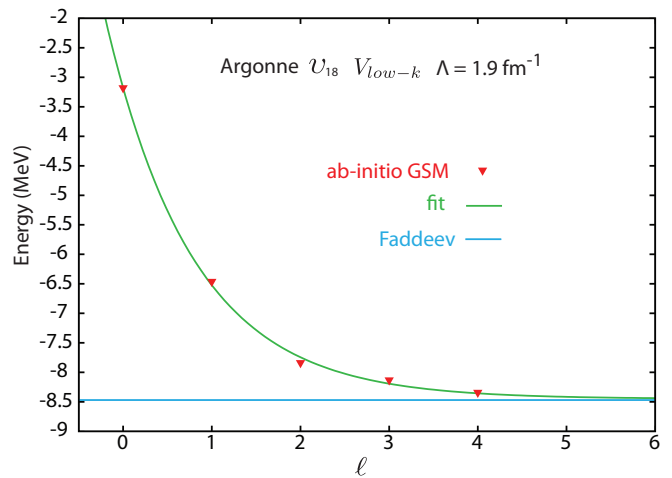}
  \caption[T]{\label{fig4}
  (Color online) $^3$H results as a function of the (maximum) angular momentum, in the interval from $\ell=0$ to $\ell=4$.  }
\end{figure}

In Fig. \ref{fig4}, we compare NCGSM results obtained with the Argonne $\upsilon_{18}$ against Faddeev calculations.
We perform a calculation using $s_{1/2}$, $p_{3/2}$, $p_{1/2}$, $d_{3/2}$, $d_{5/2}$, $f_{7/2}$, $f_{5/2}$, $g_{9/2}$, $g_{7/2}$ partial waves for protons and neutrons. The basis generating potential is the GHF which gives the 0$s_{1/2}$ proton and neutron states bound, with energies  $-10.417$ MeV and $-11.982$ MeV  along the imaginary momentum axis. For this reason, the scattering continua $i\{s_{1/2}\}, i\{p_{3/2}\},  i\{p_{1/2}\}$ are chosen along the real axis and each of them is discretized with 18 points.
Here $i$ denotes the number of discretization points which ranges from 1 to 18 for the $s$-wave and from 0 to 18 for the $p$-waves.
For the $d$-waves we choose a discrete HO basis with $b=1.5$ fm.  In particular, we use five HO states for the $d_{3/2}$ and $d_{5/2}$ states, which from now on we denote as $5d$. The physical argument behind this choice lies in the fact that for $\ell > 1$ the centrifugal barrier is large enough to confine the s.p. states and, hence, the HO basis becomes a realistic alternative. 
For $f$ and $g$ partial waves we used three HO states with $b=1.5$ fm for both protons and neutrons. In total, the s.p. basis consisted of 154 partial waves. The result we obtain is:
 \begin{equation}
E_{\mathrm{NCGSM}} = \mathrm{-8.39~MeV} \nonumber,
\end{equation}
whereas the Faddeev result \cite{nogga_scott_ach} is:
 \begin{equation}
E_{\mathrm{Fadeev}} = \mathrm{-8.47~MeV}  \nonumber.
\end{equation}
In Fig.\ref{fig4}, we show also an exponential fit of the NCGSM results for different $\ell$ of the s.p. basis. The extrapolated result is
 \begin{equation}
E_{\rm extrp} = \mathrm{-8.449 \pm 0.087~MeV} \nonumber,
\end{equation}
where the fit function is: $E = E_{\rm extrp} + b \times e^{-c \times \ell}$.
We see that inclusion of partial waves with angular momentum larger than $\ell = 4$, should have  a very small contribution, of the order of 50 keV, to the many-body result.
The g.s of $^3$H being well-bound, the set of HO shells can also provide a basis in which the many-body 
 Schr\"{o}dinger equation can be solved. But nevertheless in our case, the $\ell = 0,1$ shells are solutions of the HF potential which admits two bound $s_{1/2}$ shells and a  continuum  set of shells. What we show in Fig.\ref{fig4} serves as an illustration of the rate of convergence of the energy, for increasing the angular momentum of our basis states ,and our aim is not to deduce any physical insight by using our Berggren basis on triton (and later on $^4$He). 
For three-particles systems we can perform an exact diagonalization using the Lanczos method and, therefore, we can test the precision of the NCGSM/DMRG approach. Since details of the DMRG
\begin{figure}[b] 
  \includegraphics[width=\columnwidth]{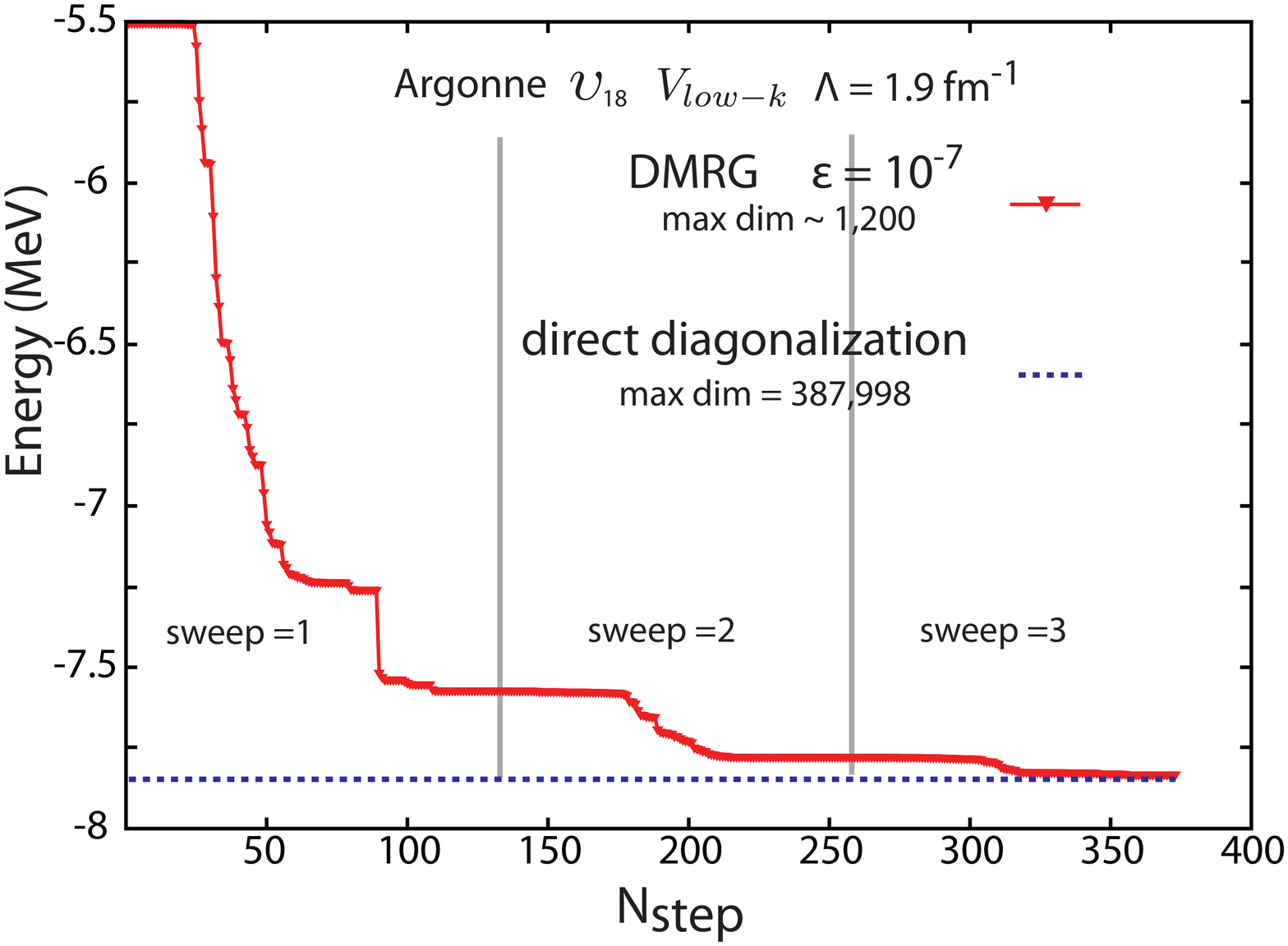}
  \caption[T]{\label{fig5}
  (Color online) Iterative process of the DMRG approach for $\epsilon$ = 10$^{-7}$ and including only waves with $\ell$ = 2 ($s-pd$). Both Lanczos and DMRG diagonalizations of the Hamiltonian are shown.}
\end{figure}
algorithm can be found in \cite{dmrg_rot2}, we only mention the basic ingredients of the DMRG calculation. 

It is important to consider in the reference space $H$, not only resonant states, but also some non-resonant shells from the $P$ space, in order to generate all possible many-body configurations in the warm-up phase, which these shells could generate \cite{dmrg_rot2}. In addition to the $0s_{1/2}$ neutron-proton resonant states, the space $H$ contained also the last $d_{5/2}$ and $d_{3/2}$
HO states. The choice of the shell to be included in the pole space $H$ is arbitrary and the results do not depend on which shell is considered. The mixture of positive and negative parity states assures that we will not miss any couplings in the
warm-up phase. In the space $P$ the shells are ordered like: $\{is_{1/2},ip_{3/2}\, ,ip_{1/2} \, ,id_{5/2}\, ,id_{3/2},\cdots \}$, where $i$ denotes the scattering shell starting from 0. In the case of the $d$-states the index i (i=0,1,...,4) denotes the HO radial quantum number.

Usually in the DMRG applications, it is decided from the beginning how many states of the density matrix that have the largest
eigenvalue will be kept. This number is then kept fixed throughout the whole iterative process. In this work we will use the truncation scheme defined in \eqref{criterion_den}. The smaller the $\epsilon$, the more vectors are kept.
Usually with $\epsilon$ = 10$^{-8}$ the exact result is reproduced. However, even for a larger $\epsilon$ (10$^{-7}$ or 10$^{-6}$) 
the agreement is very satisfactory, especially if the DMRG calculations are followed for more than two sweeps (see also Fig. \ref{fig5}). This method is called the dynamical block selection approach because the number of states $N_\rho$ kept changes during the iterative process to satisfy the truncation limit \eqref{criterion_den}.
\begin{figure*} 
  \includegraphics[width=1.4 \columnwidth]{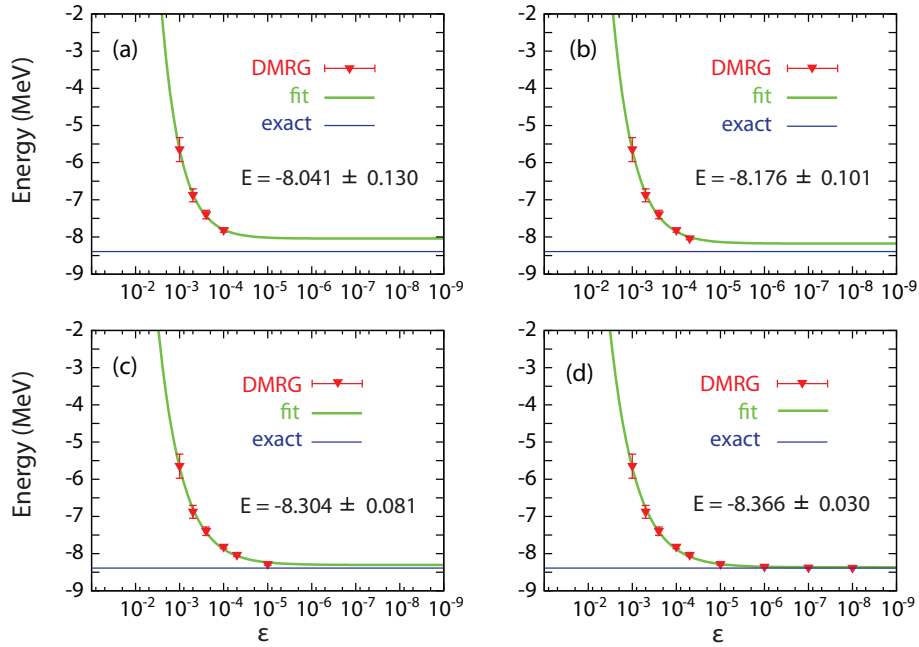}
  \caption[T]{\label{fig6}
  (Color online) Convergence of $^3$H g.s. energy as a function of the truncation error $\epsilon$ of the density matrix \eqref{criterion_den}. The s.p. basis consisted of all states up to $\ell=4$. Panel (a) corresponds to a fit with four points, panel (b) with five, panel (c) with six and panel (d) with nine points. The fit function has the form: $E = E_{\rm extrp} + b \times \epsilon^c$.  For more details, see the description in the text.}
\end{figure*}
In Fig. \ref{fig5}, we show the iterative DMRG process that includes partial waves up to $\ell$ = 2. The number of steps starts from zero, denoting the first shell in the sweep-down phase. As discussed earlier, each step corresponds to the addition of a new s.p. shell. We notice a periodic pattern of pronounced oscillations in energy, that appear in the middle of each sweep, with their amplitude continuously decreasing with the addition of more shells until the convergence is reached. Finally, the energy has converged in the end of the third sweep. The truncation error for this calculations was $\epsilon$=10$^{-7}$ resulting in a
maximum number of vectors kept $N_\rho\sim$85 and a maximum dimension $D_{\rm max}\sim$1200 of a matrix to be diagonalized.
For a direct diagonalization in this basis, the maximum dimension would be: 387,998 in $m$-scheme and 96,883 in $J$-scheme.
The exact result of the diagonalization is:
\begin{equation}
E_{\mathrm{exact}} = \mathrm{-7.840~MeV}  \nonumber,
\end{equation}
and the result obtained by the DMRG is
 \begin{equation}
E_{\mathrm{\epsilon \, = \,10^{-7}}} = \mathrm{-7.832~MeV}  \nonumber.
\end{equation}

\begin{figure*} 
  \includegraphics[width=1.4 \columnwidth]{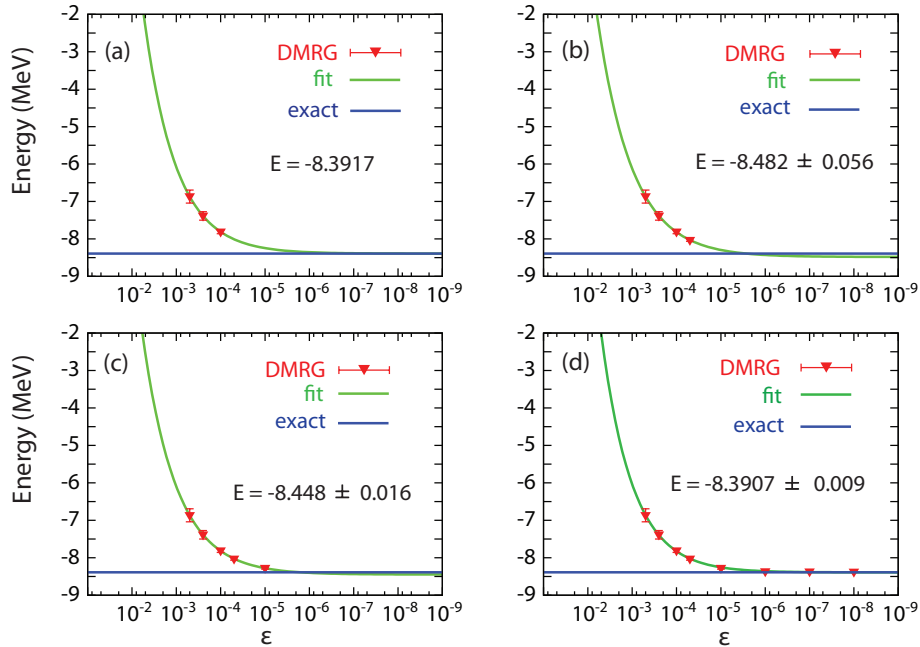}
  \caption[T]{\label{fig7}
  (Color online) Same as in Fig. \ref{fig6} but omitting the low-precision $\epsilon$ = 10$^{-3}$ point. }
\end{figure*}

\begin{figure}[h!] 
  \includegraphics[width=\columnwidth]{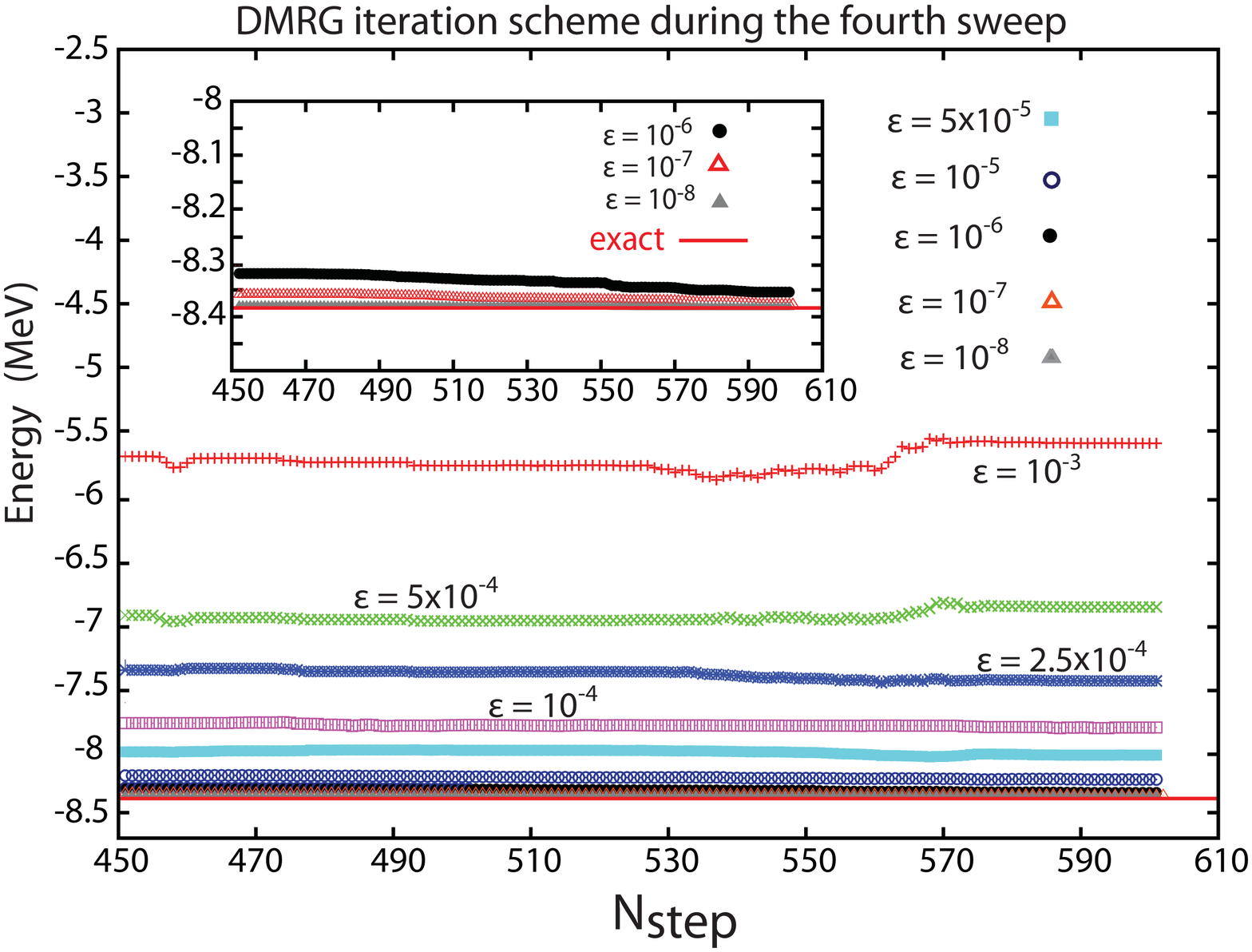}
  \caption[T]{\label{fig8}
  (Color online) Energy variation with respect to $N_{step}$ of the DMRG calculation for $^3$H and for several $\epsilon$ values. The iteration is during the fourth sweep.  }
\end{figure}

 In the following, using the V$_{low-k}$ Argonne $\upsilon_{18}$ potential, we will test the convergence of the DMRG method with respect to the truncation parameter $\epsilon$ which controls how many vectors of the density matrix are kept at each step. The results are gathered in Figs. \ref{fig6} and \ref{fig7}. Additionally, in Table \ref{tab:2} we show the number of vectors that are kept for different values of the parameter $\epsilon$ and the corresponding energy.
 \begin{table}[ht]
\caption{Vectors of the density matrix that are kept for different values of $\epsilon$. The corresponding matrix dimensions and many-body energies for $^3$H are also shown.}
\begin{center}
\begin{tabular}{cccc}
  \hline
   $\epsilon$& $\#$ vectors & dimension & energy (MeV) \\
   \hline
   10$^{-3}$ &  5   &  43     &  -5.648 \\
  5.0$\times$10$^{-4}$ &  8   &  99     &  -6.878 \\
  2.5$\times$10$^{-4}$ &  13   &  109     &  -7.396 \\
  10$^{-4}$ &  22   &  173     &  -7.821 \\
  5.0$\times$ 10$^{-5}$ &  28   &  308     &  -8.042 \\
  10$^{-5}$ &  46   &  600     &  -8.287 \\
  10$^{-6}$ &  73   &  1075     &  -8.357 \\
  10$^{-7}$ &  96   &  1909     &   -8.381 \\
   10$^{-8}$ &  117   &  2575     &   -8.388  \\
  \end{tabular}
\label{tab:2}
\end{center}
\end{table}
 Each point in Figs. \ref{fig6} and \ref{fig7} corresponds to the value at the end of the fourth sweep of the DMRG process.
 The error bar reflects the extremum values of the energy in the last sweep (see also Fig. \ref{fig8}). We observe that with decreasing the
 value of $\epsilon$ the error bar also decreases and almost vanishes for precisions better than 10$^{-5}$, {\em i.e.} the DMRG result
 quickly converges to the exact result for $\epsilon$ smaller than 10$^{-5}$.

The purpose of this exercise is to test the extrapolation properties of the NCGSM/DMRG calculations.
 In general, one would like to perform calculations with the best precision possible, {\em i.e.,} $\epsilon\sim10^{-8}$, what is presently possible only in the lightest systems. Extending the NCGSM applications to somewhat heavier systems, one needs to choose a different strategy that makes the calculation feasible and, at the same time, gives a rather precise result. The $^3$H studies serve here as a testing-ground for this investigation. 
 In Fig. \ref{fig6} we show calculations with $\epsilon$ ranging from 10$^{-3}$ up to 10$^{-8}$.
 Panel (a) consists of four points that correspond to several values of $\epsilon$ from 10$^{-3}$ up to 10$^{-4}$.
 The result is almost 600 keV away from the exact one and the extrapolation is also of a poor quality, lying
 at $\sim380$ keV away. The function we use for fitting/extrapolating numerical data has the form:
\begin{equation}
E = E_{\rm extrp} + b \times \epsilon^c \nonumber
\end{equation}
Adding one more point, the situation is improved, but still the exact result is off
 the error bars of the extrapolation (panel (b)). In panel (c) the exact result is in between the error bars of
 the extrapolation and considering all values up to 10$^{-8}$ we fall onto the exact result. 
 
 The outcome is that carrying calculations with a relatively low precision, one can find the exact result within the error bars of the extrapolated value. In this specific example, the maximum dimension for $\epsilon=10^{-5}$ (panel (c)) is $D_{\rm max}=600$ (see Table \ref{tab:2}), whereas the dimension in a direct diagonalization is 890,021 in $m$-scheme and 123,835 in $J$-scheme.

In Fig. \ref{fig7} we make the same analysis but neglecting the energy point corresponding to $\epsilon=10^{-3}$. This energy is a rather poor approximation of a final result, lying almost 3 MeV away. By extrapolating, we retrieve the exact result even in the case (panel (a)), where energies that correspond to $\epsilon$ up to 10$^{-4}$ were used in the fit. For this value of
$\epsilon$, only $N_{opt}$ = 22 vectors of the density matrix were kept, resulting in a maximum dimension of the matrix
to be diagonalized $D_{\rm max}=173$, which is much smaller than the dimension in a full diagonalization. Notice that the extrapolated value in panel (a) is not accompanied by an error estimate. This is because we are fitting here three points with 
a function that is characterized by three parameters and by definition the $\chi^2$ fit produces a zero error.

In addition, in Fig. \ref{fig8} we show the behavior of the energy during the fourth sweep in the DMRG process.
Due to the small number of vectors kept, calculations that correspond to truncation parameters 10$^{-3}$, 5.0$\times$10$^{-4}$
and to a lesser extent for 2.5$\times$10$^{-4}$, show variations with respect to the number of shells added ($N_{step}$). The error bars
in Figs. \ref{fig6} and \ref{fig7} were calculated by the extremum values of these variations. For $\epsilon$ values equal 10$^{-4}$ and smaller, the energy is a flat curve with variations of the order of less than 1 keV. 

Finally, we have calculated the expectation value of the CoM operator (Eq. \eqref{Hcm_op}). 
In the largest model space that we used, the expectation value of the H$_{cm}$ is approximately 7 keV. 
The $\hbar\omega$ energy in this case is 18.5 MeV ($b=1.5$ fm). Following the assertion of Ref. 
\cite{Gaute_Thomas_CoM}, we conclude that in a sufficiently large model space the NCGSM wave function factorizes and the CoM wave function $|\psi_{\mathrm{CoM}}\rangle$ is a Gaussian with $\hbar\omega$ = 18.5 MeV.  

The smooth convergence properties of the NCGSM/DMRG procedure both with the number of partial waves and the truncation error, as they were tested in $^3$H and $^3$He, will be used later to calculate somewhat heavier nuclei. One can perform several smaller scale NCGSM calculations, changing both the number of density matrix vectors and the number of partial waves, and extrapolate these data to  retrieve the exact result within the error bars.

\subsection{$^4$He nucleus}

In this section we apply the NCGSM to $^4$He nucleus, using the DMRG as the diagonalization technique.
\begin{figure}[b] 
  \includegraphics[width=\columnwidth]{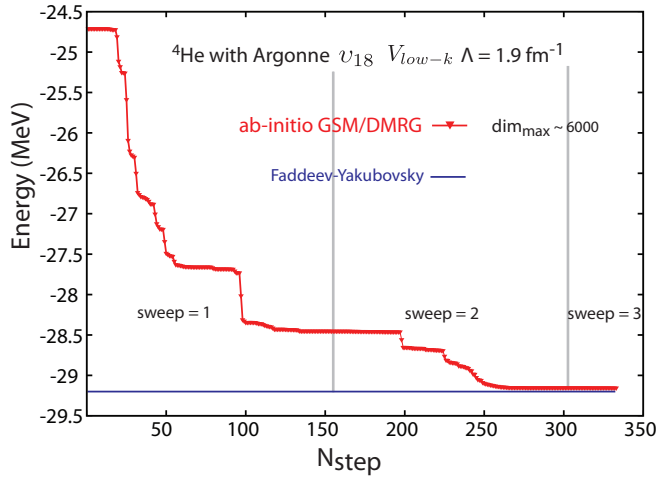}
  \caption[T]{\label{fig9}
  (Color online) Comparison of the NCGSM with the FY result with the  $V_{low-k}$ Argonne $\upsilon_{18}$ interaction. }
\end{figure}
We will compare the NCGSM/DMRG results with FY calculations, using the Argonne $\upsilon_{18}$ interaction with a $V_{low-k}$ cutoff $\Lambda$ = 1.9 fm$^{-1}$. The energies of the 0$s_{1/2}$ neutron and proton as they are calculated from the GHF process are $-26.290$ MeV and $-24.453$ MeV, respectively. As in the case of $^3$H, the $s$ and $p$ scattering continua are taken along the real axis and are discretized with eighteen points. The contours extent up to 4 fm$^{-1}$. For the remaining $d,f,g$ partial waves, we assume HO basis functions with $b=1.5$ fm.
We take $5d$, $3f$ and $3g$ states for protons and neutrons. For a truncation error $\epsilon$ = 10$^{-6}$, the maximum number of
vectors, which are kept is N$_{opt}$ = 180, resulting in a Hamiltonian matrix of dimension $D_{\rm max}\sim$6000. On the other hand, the dimension of the Hamiltonian matrix for a direct Lanczos diagonalization is computed to be 119,864,088 in $m$-scheme and 6,230,512 in $J$-scheme.

In Fig. \ref{fig9} we show the iteration pattern for $^4$He, which is similar to the one in  $^3$H case, but we observe that already in the middle of the third sweep the energy is converged. This can be attributed to the larger number of vectors of the density matrix that are kept. The converged energy is:
\begin{equation}
E_{\mathrm{NCGSM}} = -29.15~\mathrm{MeV}, \nonumber
\end{equation}
and the FY result \cite{nogga_scott_ach} is:
\begin{equation}
E_{\mathrm{FY}} = -29.19~\mathrm{MeV}, \nonumber
\end{equation}

Results of $^3$H and $^4$He using the Argonne $\upsilon_{18}$ are in a nice agreement with both FY and CC calculations with triples corrections \cite{Gaute_benchmark}. 

We also performed calculations using the chiral N$^3$LO interaction ($\Lambda$ = 1.9 fm$^{-1}$). The s.p. energies for the 0s$_{1/2}$ neutron and proton poles are $-24.333$ MeV and $-24.303$ MeV, respectively.
\begin{figure}[h!] 
  \includegraphics[width=\columnwidth]{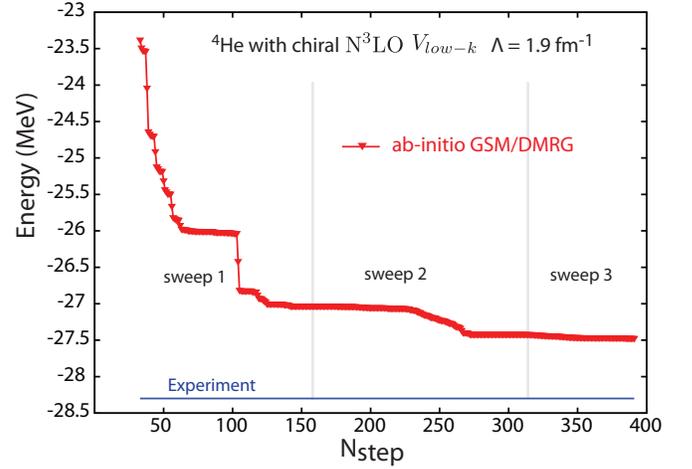}
  \caption[T]{\label{fig10}
  (Color online) Same as in Fig. \ref{fig9} but with the $V_{low-k}$ N$^3$LO interaction.  }
\end{figure}
The first point in the number of steps corresponds to approximately fortieth shell in the sweep-down phase. 
The results are shown in Fig. \ref{fig10}, and the converged energy:
\begin{equation}
E_{\mathrm{NCGSM}} = -27.48~\mathrm{MeV}. \nonumber
\end{equation}
The experimental total binding energies are found in the 2003 Atomic Mass Evaluation II (AME) \cite{Wapstra1}.
The difference with the experimental binding energy is attributed to the missing 3N interactions, which are both 'bare' and 'induced'.  As 'bare' we denote the 3N forces that appear due to the neglect of the quark degrees of freedom. The chiral potential we use has a certain cut-off, beyond which the missing physics is integrated out and this results in the appearance of many-body forces \cite{coon,u_kolck,nogga,navratil_ormand,gfmc}. As 'induced' we denote the 3N forces that are related to the renormalization technique \cite{vlowk_improv,Jurgenson,Roth}.

\subsection{$^5$He nucleus}\label{5he_chap}

\begin{figure}[b] 
  \includegraphics[width=\columnwidth]{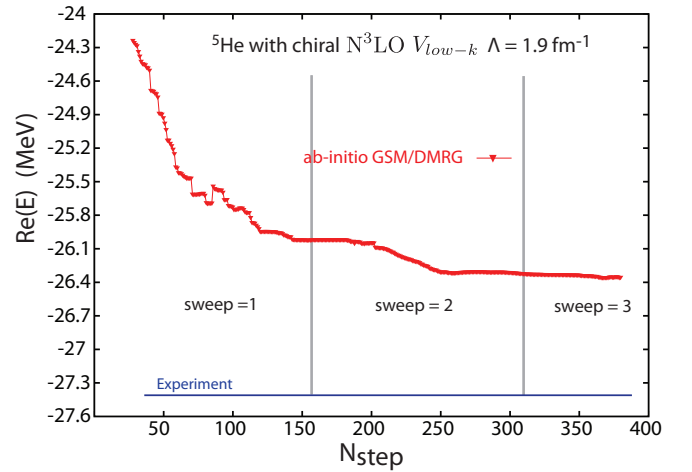}
  \caption[T]{\label{fig11}
  (Color online) Same as in Fig. \ref{fig10} but for the real part of $^5$He g.s. energy.  }
\end{figure}

$^5$He is a challenge for any many-body theory due to its unbound character. In particular, both the ground and first excited states are many-body resonances, which obey outgoing asymptotics. Because of these characteristics, the complex energy formulation of the NCGSM using the Berggren ensemble is suitable for its description. 
Indeed, in our formalism the resonance parameters (g.s energy with respect to $^4$He and the width) will be identified as the eigenvalues of the complex-symmetric Hamiltonian matrix. 
The position of the resonance will then be the real part of the energy, while the imaginary part is related to the
width by the formula: $\Gamma  $ = -2Im(E). An advantage of using Berggren's states is that the complex
eigenvalues (E) are the ones, which correspond to states with purely outgoing-wave solutions. 
Contrary to the previous applications, where we used a real-energy basis, in the case of $^5$He we are employing a complex basis, which also includes the 0$p_{3/2}$ resonant state. Overall we include the bound 0$s_{1/2}$ neutron state with a s.p. energy of $-23.290$ MeV, the bound 0$s_{1/2}$ proton state with s.p. energy $-23.999$ MeV and the 0$p_{3/2}$ s.p. resonance with a real part of energy 1.193 MeV and a width 1267 keV. Its position in the complex-energy plane is:  $k = (0.277,-0.068)$ fm$^{-1}$. 
The s.p. basis for protons and neutrons is produced by a GHF calculation using the N$^3$LO $V_{low-k}$ interaction with  $\Lambda$ = 1.9 fm$^{-1}$. The $p_{3/2}$ contour is taken complex to satisfy the Berggren completeness relation \eqref{complet}, whereas the $s_{1/2}$ and $p_{1/2}$ contours may be chosen along the real-$k$ axis. For states with $\ell > $1, we assume the HO basis functions ($5d,3f,3g$) as we described in the previous cases. 

In Fig. \ref{fig11} we show the DMRG convergence pattern of the real part of the g.s. energy in $^5$He. The calculation in Fig. \ref{fig11} is presented starting from the  fortieth  shell in the sweep-down phase. The converged energy:
\begin{equation}
\Re{\rm e}\left(E_{\rm NCGSM}\right) = -26.31~\mathrm{MeV}, \nonumber
\end{equation}
lies at about 1 MeV above the experimental total binding energy \cite{Wapstra1}. The truncation error in this calculation is $\epsilon$ = 10$^{-6}$ and the maximum number of vectors we kept is $N_{opt}$ $\sim$ 300. The corresponding dimension of the matrix is $D_{\rm max}\sim$10$^5$, whereas in the direct diagonalization one deals with a matrix of a dimension $\sim$3$\times$10$^{9}$. 

\begin{figure}[t] 
  \includegraphics[width=\columnwidth]{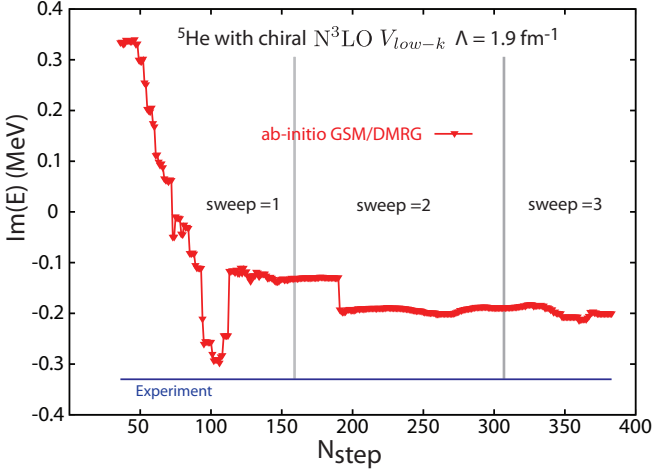}
  \caption[T]{\label{fig12}
  (Color online) Same as in Fig. \ref{fig11} but for the imaginary part of $^5$He g.s. energy.  }
\end{figure}

The imaginary part of the $^5$He g.s. energy is shown in Fig. \ref{fig12}. 
The converged value is:
\begin{equation}
\Im{\rm m}\left(E_{\rm NCGSM}\right) = -0.2~\mathrm{MeV}, \nonumber
\end{equation}
\textit{i.e.}, $\Gamma_{\rm NCGSM}$ = 400 keV.
Having calculated the g.s. total binding energies of $^4$He and $^5$He, we obtain then the position of the resonance at an energy 1.17 MeV above the
$\alpha$ + n threshold with a width $\Gamma$ = 400 keV. 

We will compare this result with the value extracted from experiment.
Experimentally, resonance parameters can be extracted in two ways \cite{Tilley}.
First, one could apply the conventional R-matrix approach on the real axis and use the Lane and Thomas prescription for the extraction of positions and widths of a state \cite{LThomas}.
The second way is the extended R-matrix approach \cite{Hale,Csoto}, which associates the resonance parameters with the complex poles of the S-matrix.
According to \cite{Tilley,private_communication}, this is the recommended prescription, and it is also the most appropriate for comparison to the NCGSM results, since our
complex eigenvalues correspond, indeed, to the location of the poles of the S-matrix, with the correct asymptotic behavior. The extended R-matrix approach gives the position
of the $(3/2)^-$ state at an energy 798 keV above the $\alpha$ + n threshold and with a width $\Gamma$ = 648 keV.
The difference we observe between our results and the experimental results is attributed to the specific interaction that we employ and/or to missing many-body forces. We know already that
the position of such unbound states, which have a non Breit-Wigner character (broad states) and  are close to threshold, heavily depends on the specific characteristics of the NN (or NNN) interaction. 
This was shown for calculations of $^5$He nucleus, in the many-body frameworks of
the NCSM merged with the Resonating Group Method (RGM) \cite{quaglioni,quaglioni2,quaglioni3}, (see also \cite{hupin} for a recent application of the NCSM/RGM with 3N forces) and also in the Green's Function Monte Carlo approach \cite{Nollett}, with
a V$_{low-k}$ $\Lambda$ = 2.1 fm$^{-1}$ Argonne $\upsilon_{18}$ and a SRG $\lambda$ = 2.02 fm$^{-1}$ N$^{3}$LO for the NCSM/RGM and a "bare" Argonne $\upsilon_{18}$ with Urbana/Illinois 3N forces in the GFMC case. 

We verify this assertion also in our NCGSM calculations. For that purpose,
we have compared NCGSM results for two different renormalization scale parameters:  $\Lambda$ = 1.9 fm$^{-1}$ and 2.1 fm$^{-1}$.
To this end we have performed the NCGSM calculation using the Davidson (variant of Lanczos) method for the diagonalization of the many-body Hamiltonian \cite{review_GSM} and truncating the space of configurations to up to 4p-4h excitations. In addition, we have used a smaller s.p. basis by taking only the neutron $p_{3/2}$ states as Berggren states, whereas remaining states up to $\ell=4$ are the HO basis functions with a HO length $b=1.5$ fm. 
The many-body energy and width of $^4$He/$^5$He in this NCGSM/4p4h calculation are close to the exact NCGSM/DMRG results. Using the V$_{low-k}$  N$^3$LO force with $\Lambda$ = 1.9 fm$^{-1}$, we obtain $E_{\rm NCGSM}^{(4p4h)} = -27.386$ MeV for the total binding $^4$He, and $\Re{\rm e}\left(E_{\rm NCGSM}^{(4p4h)}\right) = -25.825$ MeV for $^5$He with a width 
$\Gamma_{\rm NCGSM}^{(4p4h)} = 370$ keV.
The $^5$He g.s. is found at an energy 1.56 MeV above the $\alpha$+n threshold, with a width $\Gamma$ = 370 keV.
In the case of $\Lambda$ = 2.1 fm$^{-1}$ renormalization parameter the total binding energy is $E_{\rm NCGSM}^{(4p4h)} = -26.060$ MeV for $^4$He and $\Re{\rm e}\left(E_{\rm NCGSM}^{(4p4h)}\right) = -23.903$ MeV for $^5$He with a width $\Gamma$ = 591 keV. Hence, for $\Lambda$ = 2.1 fm$^{-1}$ the g.s. of $^5$He is found at a position 2.15 MeV above the $\alpha$ + n
threshold and with a width $\Gamma$ = 591 keV. As expected, we observe that the unbound state of $^5$He is very sensitive on the choice of the NN interaction.

It needs to be mentioned that, because of the broad nature of the $^5$He g.s. and the proximity to the threshold, results would also depend on the method used to extract such a state from the experimental data. 
We find that if one uses the conventional R-matrix approach on the real axis, the position of the g.s. is
at 0.963 MeV above the $\alpha$ + n threshold, with a width $\Gamma$ = 985 MeV and both values are different by about 200 keV as compared to the extended R-matrix estimations.
The R-matrix estimation of \cite{bond_firk} is similar to the extended R-matrix method, since the authors determined the resonance energy and width by 
finding the S-matrix pole location implied by the fit parameters.
In Table.\ref{tab:3} we gather the existing measurements and/or the extracted values from the data of the position and width of $^5$He g.s. 
\begin{table}[ht]
\caption{NCGSM$_{\rm DMRG}$ result as compared to experimental position and width of the $^5$He g.s. Energies are with respect to the $\alpha$ + n threshold.}
\begin{center}
\begin{tabular}{ccc}
  \hline
   Method &  Energy (MeV) & $\Gamma$ (MeV)\\
   \hline
  NCGSM$_{\rm DMRG}$ &  1.17   &  0.400   \\
  extended R-matrix \cite{Tilley} &  0.798   &  0.648     \\
  R-matrix \cite{Tilley} &  0.963   &  0.985     \\
  R-matrix \cite{bond_firk} & 0.771 & 0.644 \\
   NUBASE evaluation \cite{Wapstra2} \footnote{In the data evaluation of \cite{Wapstra2} the half-life T$_{1/2}$ was given to be 700 ys, where ys stands for yoctosecond and equals 10$^{-24}$ s.
   Then the width was obtained by the relation: $\Gamma_{\rm cm}$ T$_{1/2}$ $\approx$ $\hbar$ln2, and $\Gamma_{\rm cm}$ is the level total width. }   &  0.890   &  0.651     \\
   $^3$He + t \cite{Smith}  &  0.79   &  0.525     \\
  \end{tabular}
\label{tab:3}
\end{center}
\end{table}

\subsection{Asymptotic normalization coefficient of a complex-energy ground state of $^5${\rm He}}

Asymptotic normalization coefficients (ANCs), as the name implies, define the overall normalization of the tail of the overlap function between a system with A and  A $\pm$ N nucleons.
In our case, we are interested in the addition of a neutron to $^4$He, so A = 4 and N = +1. Because of their definition, they have attracted a lot of attention, since the
properties of the wavefunction tail can be useful to test many-body methods. Additionally, their importance is manifested in the borderline between nuclear physics and
astrophysics, where for example their knowledge is necessary for the correct description of neutrinos reaction rates \cite{Akram_exp,2006_exp}.
Contrary to Spectroscopic Factors (SFs), ANCs are observables, due to their invariance under short-range unitary transformations.
This can be understood intuitively because of their connection to the tail of the overlap function. Indeed, for a truly unitary transformation (such as V$_{low-k}$ or SRG),
where generated many-body terms will not be neglected, what will change is the short range part of the potential and the interior region of the wavefunction. The tails of both potential and wavefunction, however, will
remain unchanged. Hence, the ANC is a quantity that does not "feel" the changes of the short range physics (see \cite{Akram}, discussion in \cite{Dick_Achim_2005} about observables and
non-observables and also \cite{vlowk_improv} for an overview of unitary transformations and their impact on nuclear structure operators.). Experimentally, ANCs have been measured for s-shell and p-shell nuclei and for a detailed list of references we refer the reader to \cite{Nollett_ancs_11}.
Recently, theoretical calculations of ANCs have been performed in the framework of the HH approach \cite{Viviani}, the GSM and SMEC \cite{ANC_pap}, the GFMC \cite{Nollett_ancs_11,brida,Nollett_12} and also \cite{Akram_mirror}.

One of the criteria that should be met in order to calculate in a meaningful manner the ANC, is the correct asymptotic behavior of
wavefunction. In the NCGSM, using the Berggren ensemble as basis, this condition is met for either bound, weakly bound or unbound states. 
The other condition is related to the separation energy S$_{\rm 1n}$. Indeed, the tail of the
overlap is very sensitive on the S$_{\rm 1n}$ and small differences in the S$_{\rm 1n}$ for different Hamiltonians can cause large differences on the ANCs. This can
be especially complicated for \textit{ab initio} calculations, where there are
no adjustable parameters. Recently, a method was
proposed to alleviate this problem in the GFMC approach, where calculations could be performed using experimental separation energies \cite{Nollett_ancs_11,Nollett_12}.

Our approach to calculate the ANC of the reaction $^4$He + n $\to$ $^5$He is based on the calculation of the overlap: 
\begin{equation}
\label{overlap_int}
\mathrm{I_{\ell j}(r) = \frac{1}{\sqrt{2J_{a} + 1}} \sum_{\mathcal{B}} \langle \widetilde{\Psi_{A}^{J_{A}}} || a^{\dagger}_{\ell j}(\mathcal{B}) || \Psi_{A-1}^{J_{A-1}} \rangle \langle r\ell j|u_{\mathcal{B}} \rangle } 
\end{equation}
where the sum runs over the complete set of basis states ${\mathcal{B}}$.
In our case, a$^{\dagger}_{\ell j}(\mathcal{B})$ creates neutron in any $p_{3/2}$ s.p. state of the Berggren basis and $ \langle r\ell j|u_{\mathcal{B}} \rangle$ is the p$_{3/2}$ radial wave function, so the radial overlap integral takes the form:
\begin{widetext}
\begin{equation}
\label{our_case}
\mathrm{I_{\ell = 1, j = 3/2}(r) = \frac{1}{2} \sum_{n} \langle \widetilde{^5He_{3/2^-}} || a^{\dagger}_{n,\ell=1,j=3/2} || ^4He_{0^+} \rangle  u_{n,\ell=1,j=3/2}(r)}
\end{equation}
\end{widetext}
Regarding the details of the calculation, we used the Davidson method to diagonalize
the Hamiltonian with a V$_{low-k}$ $\Lambda$ = 1.9 fm$^{-1}$ potential and the model space was the same as the one at the end of Sect. \ref{5he_chap}.
We also performed calculations at a cut-off $\Lambda$ = 2.1 fm$^{-1}$.
  
\begin{figure}[t] 
  \includegraphics[width=\columnwidth]{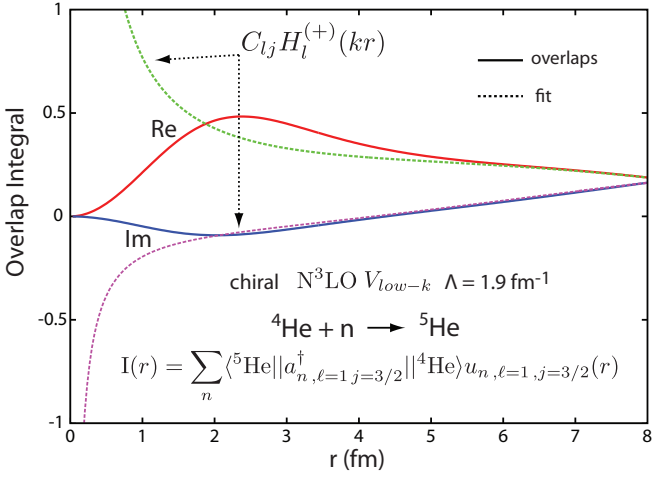}
  \caption[T]{\label{fig13}
  (Color online) The radial overlap integral for $^4$He$_{0^+}$ + n $\to$ $^5$He$_{3/2^-}$. The asymptotic region is fitted by the Hankel function. See the text for more details. }
\end{figure}
\begin{figure}[t] 
  \includegraphics[width=\columnwidth]{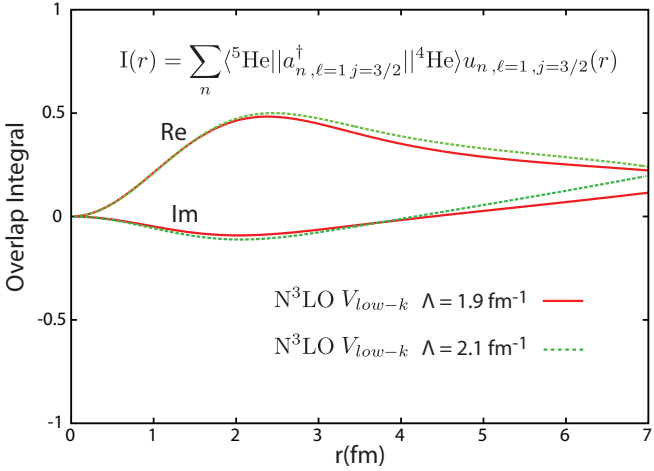}
  \caption[T]{\label{fig14}
  (Color online) Real and imaginary parts of radial overlap integrals for two different V$_{low-k}$ cutoff scales 
  $\Lambda$. }
\end{figure}
The radial overlap integral for $^4$He$_{0^+}$ + n $\to$ $^5$He$_{3/2^-}$ is shown in Fig. \ref{fig13} for $\Lambda$ = 1.9 fm$^{-1}$. By fitting real and imaginary parts of the radial overlap integral in the asymptotic region with a Hankel function, one can extract the ANC, which in this case equals 0.197 fm$^{-1/2}$, whereas for $\Lambda$ = 2.1 fm$^{-1}$ equals 0.255 fm$^{-1/2}$.
Real and imaginary parts of the radial overlap integrals for $\Lambda$ = 1.9 fm$^{-1}$ and 2.1 fm$^{-1}$ are compared in Fig. \ref{fig14}.  
We immediately see that the two different Hamiltonians we employed produced different separation energies, the overlaps have different asymptotic behavior
and the ANCs are different. Thus, ANC calculations
are meaningful only in the case that the separation energy is under
control, which is not true in our case.  Of course, the reason for this
behavior is that the scheme we used to alter the short-range physics
is not a unitary transformation, and additionally, we neglected a "bare" 3N interaction before 
renormalization.  Unitarity was violated by the neglect
of many-body forces, which yield different separation energies and ANCs.
Nevertheless, calculations of ANCs can be used as a consistency test
for the many-body method proposed, as we will see in the following
discussion.
Indeed, the width and ANC are related via the relation (see Ref. \cite{ANC_pap,Nollett_12} and references cited therein)
\begin{equation}
\label{extraction}
C = \sqrt{\frac{\Gamma \mu}{\hbar^2 \Re{\rm e}{(k)}}  }
\end{equation}
which is mainly valid for narrow resonances, even though in our case we are describing a relatively broad state. This formula expresses actually the outgoing flux probability for the
decaying (outgoing-wave)  $^5$He g.s. 
In this relation, $C$ denotes the ANC, $\mu$ is the effective mass (1.25 in our case) and $k$ is the linear momentum associated with the one-neutron separation energy of $^5$He. 
We then find a width $\Gamma_{\mathrm{ANC}}$ = 311 keV and  $\Gamma_{\mathrm{ANC}}$ = 570 for $\Lambda$ = 1.9 fm$^{-1}$ and 2.1 fm$^{-1}$ respectively, which is consistent with the width obtained in NCGSM/DMRG (or NCGSM/4p4h). As we said, relation \eqref{extraction} is mainly valid for narrow resonances, which is known for calculations on the real-axis \cite{Akram_Tribble} and it was also proved in \cite{ANC_pap} for an unbound state in the complex plane.
The approximation is that the real part of the linear momentum $k$ = $\sqrt{\frac{2m\rm S_{\mathrm{1n}}}{\hbar^2}}$ is considered. However, for a complex state as in the case of $^5$He, we have:
\begin{equation}
k  \sim \sqrt{ \rm S_{1n} ( 1 - i\frac{\Gamma}{2\rm S_{1n}} )   },
\end{equation}
where S$_{\rm 1n}$ stands for the one-neutron separation energy.
The condition then, for a real number linear momentum, is that of
\begin{equation}
\frac{\Gamma}{\rm 2S_{1n}} \, \to  0.
\end{equation}
We see that the notion of a narrow resonance, does not imply $\Gamma$ $\to$ 0, but it is actually the value the width $\Gamma$ has, with respect to the separation energy, that has to be small. 
In our calculations, this quantity is in the range of about 10$\%$ to 15$\%$ for the two renormalization parameters we considered and this error is compatible with the comparison
between the width from the NCGSM diagonalization and its extraction from the ANC formula \eqref{extraction}.

\subsubsection{$p_{3/2}$ spectroscopic factor in the ground state of $^5${\rm He}}

Spectroscopic factors are solely a theoretical invention and should not be interpreted as a measurable quantity.
The spectroscopic factor is given by the real part of the norm $S^2$ of the radial overlap integral: 
\begin{equation}
\label {SF}
S^2 = \int\hspace{-1.4em}\sum_{\mathcal{B}} \langle \widetilde{\Psi^{J_{A}}_{A}} || a^+_{\ell j}(\mathcal{B}) || \Psi^{J_{A-1}}_{A-1} \rangle^2
\end{equation}
The overlap is always a representation-dependent quantity, since it involves the interior of the wavefunction, which will change
accordingly depending on the scheme the practitioner will use. As scheme or representation we mean here the renormalization of the short-range physics, namely the NN potential.
Indeed, one has the freedom to change the off-shell behavior of the potential in any possible manner, while maintaining the NN phase-shifts (phase-shift equivalent potentials).
Each method of changing the off-shell behavior is then considered as a different representation or scheme.
Nevertheless, SFs can provide information on shell occupancies and they can be a measure of correlations, i.e. how much from
the s.p. picture the nucleus deviates. Such overlaps have been calculated by several methods, either in \textit{ab initio} or more phenomenological approaches \cite{peter_carlos,peter_2004,Viviani,brida,Nollett_ancs_11,ANL_overlap_list,Timof,Oivind,ANC_pap,rf:18a} (see also discussion on ANCs).

Using NCGSM/4p4h solutions for $^4$He/$^5$He, one finds that the spectroscopic amplitude of $^5$He corresponding to the channel $\left[ {^4}{\rm He}({\rm g.s.})\otimes p_{3/2} \right]^{3/2^-}$ is 0.787 ($S^2$ = 0.62) for $\Lambda$ = 1.9 fm$^{-1}$ and 0.812 ($S^2$ = 0.66) for $\Lambda$ = 2.1 fm$^{-1}$.
At this point we perform a calculation at a cut-off scale $\Lambda$ = 1.5 fm$^{-1}$. The total binding for $^4$He is -28.670 MeV and for $^5$He -27.285 MeV.
We gather our numbers for the three different cut-offs on Table \ref{tab:4}.
\begin{table}[ht]
\caption{NCGSM$_{\rm (4p4h)}$ results for the $S$ dependence on the V$_{low-k}$ cut-off $\Lambda$ with respect to the S$_{1n}$.}
\begin{center}
\begin{tabular}{llc}
  \hline
   $\Lambda$ fm$^{-1}$ &  S$_{1n}$ (MeV) & $S$ \\
   \hline
  2.1 &   -2.15  &  0.812   \\
  1.9 &  -1.56  &  0.787    \\
  1.5 &   -1.38   &   0.774   \\
  \end{tabular}
\label{tab:4}
\end{center}
\end{table}
The SF in the case of $\Lambda$=1.5 fm$^{-1}$ is also, slightly, reduced. Overall, the SF is reduced when the separation energy approaches the threshold. This behavior confirms the findings of GSM calculations on the anomalous
behavior of SFs, for close to threshold states \cite{rf:18a}.  Such  a quenching of SFs was found in Coupled-Cluster calculations \cite{Oivind} and also in \cite{Timof}.

\section{Conclusions and future perspectives}
\label{conclusions}

In this work, we applied the Berggren s.p. ensemble to perform \textit{ab initio} NCGSM/DMRG calculations for selected light nuclei, both well bound and unbound. We used a translational invariant Hamiltonian and benchmarked our results against Faddeev and Faddeev-Yakubovsky calculations for $^3$H and $^4$He, respectively. We also investigated the extrapolation properties of the NCGSM/DMRG iterative procedure with the number of partial waves and the truncation error. We found that even if a relatively small number of vectors of the density matrix are kept, the NCGSM results can be extrapolated with high accuracy to the exact result. This methodology will be followed for heavier systems, where the matrix dimensions, even with the DMRG algorithm cannot be handled at present. 

The NCGSM is a  natural choice to calculate unbound nuclei, such as $^5$He. For the description of $^5$He,
we employed a complex-energy Berggren basis consisting of bound 0$s_{1/2}$ proton/neutron s.p. states, the 0$p_{3/2}$ neutron s.p. resonance, and the associated real and complex non-resonant continua. We successfully reproduced the unbound character of this system from first principles using the N$^3$LO chiral potential as the NN interaction. For  $\Lambda$ = 1.9 fm$^{-1}$, the calculated neutron separation energy and the neutron emission width of the $^5$He g.s. are in a reasonable agreement with the experimental data. Still the NCGSM binding energies of $^4$He and $^5$He for this interaction are less by about 1.5 MeV than their experimental values. 

Truncating the NCGSM configuration space up to 4p-4h excitations, we were also able to calculate the radial overlap integral, the spectroscopic factor and the ANC for $^4$He${_{0^+}}$ + n $\to$ $^5$He${_{3/2^-}}$. The one-neutron emission width of $^5$He associated with this ANC was found to be in agreement with the width obtained by the many-body diagonalization, which is a nice consistency test for our method. 

Quantities that are related to the "tail" of the wavefunction, such as the ANC and the width, can be sensitive to modifications of the
short-range part of the interaction, if the latter is not done in a consistent (i.e. preserving the unitarity) manner. Within the NCGSM we can probe the impact of missing many-body terms, which
arise from the renormalization of the short-range physics, or the impact of neglecting many-body terms from the generic nuclear Hamiltonian, on quantities which are characteristic
of unbound systems (i.e. widths) and /or quantities such as the ANCs, which are relevant in regions outside the nuclear attraction where the correct asymptotic behavior is important. 

The correct asymptotic behavior of the system and the coupling to the continuum plays an important role in the reaction theory \cite{baroni}. The GSM has been recently generalized for the study of reactions within a Coupled Channel (CC) GSM framework \cite{yan02}. In this respect, the NCGSM can provide the realistic wave function for a target nucleus, which will include both many-body correlations and coupling to the continuum.

This work serves as a proof of principle of the application of the Berggren's basis in a NCSM framework.
In the near future, we plan to apply the NCGSM supplemented with the DMRG iterative procedure to calculate excited states of $^4$He and $^5$He, heavier weakly bound systems, such as $^6$He, and very exotic systems in the hydrogen isotopic chain. Understanding the role of
three-nucleon forces and continuum coupling in light nuclei at the limits
of nuclear stability will be important challenges for future NCGSM studies.

\begin{acknowledgements}
Discussions with G.M. Hale and G. Hagen are gratefully acknowledged. This research was supported by an allocation of advanced computing resources provided by the National Science Foundation. The computations were performed on Kraken at the National Institute for Computational Sciences (http://www.nics.tennessee.edu/). This research used resources of the Oak Ridge Leadership Computing Facility at the Oak Ridge National Laboratory, which is supported by the Office of Science of the U.S. Department of Energy under Contract No. DE-AC05-00OR22725.
 This work was supported through FUSTIPEN (French-U.S. Theory Institute for Physics with Exotic Nuclei) under DOE grant number DE-FG02-10ER41700. Partial support of this work is also acknowledged by B.R.B. and G.P. through NSF grant number PHY-0854912 and by J.R. through the European Research Council
under the European Community's Seventh Framework Programme (EP7/2007-2013)/ERC grant agreement no.240603. 

\end{acknowledgements}

\nocite{*}
\bibliographystyle{unsrt}
\bibliography{ab_initioGSM_prev1}    

\end{document}